\documentstyle[12pt,draft,lscape,array,nature12]{article}
\def\beq{\begin{equation}}
\def\eeq{\end{equation}}
\def\bey{\begin{eqnarray}}
\def\eey{\end{eqnarray}}
\def\pc{\, {\rm pc} }
\def\kpc{\, {\rm kpc} }
\def\msun{M_\odot}

\begin{document}
\title{Earth-mass dark halos are torn into dark mini-streams by stars}
\author{HongSheng Zhao$^1$, 
\thanks{$^1$PPARC Advanced Fellowship at University of St Andrew's, KY16 9SS, UK. 
Young Overseas Scholar at Beijing Observatory.  E-mail: hz4@st-and.ac.uk}
James Taylor$^2$, 
\thanks{$^2$Astronomy Department, Caltech, mail code 105-24, 1200 East California, Pasadena CA 91125,
Email: jet@astro.caltech.edu}
Joseph Silk$^3$ \& Dan Hooper$^3$
\thanks{$^3$Astrophysics, University of Oxford, Denys Wilkinson Building, Keble Road, Oxford OX1 3RH
Email: silk@astro.ox.ac.uk, hooper@astro.ox.ac.uk}
}
\date{Communications Arising from {\it Letter to Nature} by Diemand et al.}
\maketitle
{\sf 
The promising neutralino dark matter particles generically 
condense into numerous earth-mass dark halos with smaller substructures
suppressed by free streaming\cite{berezinsky}.
The recent {\it Letter to Nature}\cite{moore} claims that 
these 0.01pc-sized dense halos emerged at redshifts 60--26 are rarely destructed inside galaxies, 
hence the nearest halos at about 0.1pc from Earth are bright in annihilation-powered gamma-rays,
but are inaccessible to direct detections. 
However, most mini-halos reaching solar neighbourhood should experience 
strong impulses by individual stars in the Galactic disk and bulge, and have been torn
into pc-long tidal streams over a Hubble time with only modest overdensity, 
reducing indirect annihilation signals. Sweeping across the solar system per century, 
mini-streams leave directional and temporal signatures for direct searches of neutralinos.
}

When a minihalo of a mean density $\rho \equiv {3 m(r) \over 4\pi r^3}$ inside radius $r$
encounters a star of mass $M_* \gg m(r)$ 
with a high relative velocity $V_r$ and impact parameter $b>r$, 
the minihalo is stripped outside a radius $r$ if the specific heating 
${\left(\delta v\right)^2 \over 2} \times {2 \over 3}$ (a projection factor
of the differential impulse $\delta v$ perpendicular to and along 
the star's almost straight path)\cite{BT}
is greater than the specific binding energy ${G m(r) \over 2r}$ at radius $r$, 
where $\delta v$ is the tidal force ${GM_* r \over b^3}$ times the effective 
duration of the impulse ${2b \over V_r}$.  Hence only the dense part of 
the minihalo survive with $G\rho > k \left({GM_* \over S}\right)^2$ where 
the factor $k=2\pi$, and $S \equiv (\pi b^2) V_r$
is the volume plunged through by the minihalo per unit time.  
We esimate $S \le H_0 \left<n_*\right>$ by requiring fewer than one disruptive encounter 
per Hubble time $H_0^{-1}$ in a sea of normal stars of mass $M_*$ with a time and 
space averaged star density $\left<n_*\right>M_*$, which is estimated as\cite{robin} 
either $\sim 0.001\msun\pc^{-3}\left({16\kpc \over 2H}\right)$ or 
$\sim 0.012\msun\pc^{-3}\left({16\kpc \over 2H}\right) 
\left({8\kpc \over R}\right)^2$  
by averaging either an old stellar local disk density $\sim 20\msun\pc^{-2}$  
over a column of orbital height $2H$,  
or an old stellar disk and bulge mass $4\times 10^{10}\msun$ 
over a cylinder of orbital radius $R$ and height $2H$.  
The survival minihalos are regions of overdensity above 
${\rho_{\rm surv} \over \rho_{\rm crit}(z=60)}=
200\left[{\left<n_*\right>M_* \over 0.001\msun\pc^{-3}}\right]^2$
where we take the prefactor $k \sim 10$
(\cite{bahcall} found $k \sim 30$ in binary disruption 
simulations including head-on encounters and 
a series of gradual heating with large impact parameters).
Scaled to the background density $\rho_{\rm crit}$ at the emerging redshift $z=60$, 
the $10^{-6}\msun$ minihalos proposed by\cite{moore} have low overdensities 
$\sim$ 4 and 200 inside half-mass radius 0.01pc and their 0.001pc resolution
limit (enclosing of order 5 percent of the minihalo mass, cf. their Fig.2) respectively,
hence these minihalos should be mostly stripped by encounters long ago.
The resupply of minihalos is insignificant locally since  
the local overdensity collapsed mainly at high redshift.

More sophisticated semi-analytical models of the massloss\cite{babul}
also predicts that mini-halos on polar orbits would lose a few percent of the mass
per orbit if they spend a few percent of the orbital period in strong tide regime 
near disk stars, hence over a Hubble time ($\sim$ 20--70 orbits at the solar radius),
a minihalo with 1m/s escape velocity is stripped nearly completely and stretched hundred fold 
to a modest overdensity region of several pc long.  
These dark streams roughly fills the local volume.  
The Earth is likely crossing one of these 0.02pc wide mini-streams on time scales of centuries. 
Sensitive, long-term direct detection experiments 
might expect to see slow variations of order 1\% per year. 
Finally, the fact that local minihalos on planar orbits with small pericentres 
are preferentially destructed could imprint an interesting directional or phase 
modulation on direct detection signals.  
While the smoothing of the local halo increases the chance of direct
detection, it simultaneously reduces the indirect signals of annihilation 
\cite{Taylor:2002zd}, which depend on neutralino density squared, hence is 
proportional to the fraction of dark substructures in the local density
and their contrast to the smooth component.
The boost factor drops from $\sim 15-45$ for unstripped concentrated substructures 
to $\sim 4-12$ if most minihalos are disrupted and stretched, although 
this boost is still larger than previous estimates \cite{berezinsky}.

\label{lastpage}

\end{document}